\documentclass{aa}  

\usepackage{graphicx}   
\usepackage{txfonts}

\usepackage[separate-uncertainty,multi-part-units = single]{siunitx}

\usepackage{hyperref}
\hypersetup{colorlinks, linkcolor={red}, citecolor={blue}, urlcolor={magenta}}

\defcitealias{2015ApJ...800L..35A}{AP15} 

\begin{document}

\title{Probing the broad line region geometry and size of the gravitationally lensed quasar Q2237+0305 with microlensing time series}
\titlerunning{Microlensing times series in Q2237+0305}
   \author{Đ.\,V.\,Savić \inst{1,2}
          \and
          D.\,Hutsemékers\inst{1}
          \and
          D.\,Sluse\inst{1}
          }

  \institute{
          Institut d’Astrophysique et de Géophysique, Université de Liège, Allée du 6 Août 19c, 4000 Liège, Belgium. \\
          \email{dsavic@uliege.be}
               \and
          Astronomical Observatory Belgrade, Volgina 7, 11060 Belgrade, Serbia
          }
  \date{Received \today;}



\abstract{Lensed quasars are powerful cosmic laboratories; they are used to  simultaneously probe various astrophysical phenomena. Microlensing by stars within distant galaxies acts as strong gravitational lenses of multiply imaged quasars, and provides a unique and direct measurement of the lensed quasar internal structure. Microlensing of the continuum emitting region as well as the broad-line region (BLR) is well characterized by four observable indices, $\mu^{cont}$, $\mu^{BLR}$, $WCI$ (wing-core), and $RBI$ (red-blue), measured directly from the spectra. During the 2004--2007 monitoring period, image A of the quadruply lensed system Q2237+0305  underwent a strong microlensing amplification, while image D remained unaffected. We used 35 epochs of archival spectrophotometric data of Q2237+0305 obtained with the  Very Large Telescope of the European Southern Observatory to develop an independent microlensing method for estimating the geometry and size of the BLR. We measured the index time series for the \ion{C}{iv} line and the continuum emission at \SI{1450}{\ensuremath{\mathrm{\mbox{\AA}}}}. We built a library of the simulated microlensing index time series that reproduce the observed times series based on three representative BLR models: Keplerian disk (KD), polar wind (PW), and equatorial wind (EW). After sampling the model parameter space, we find that KD is the predominant model, while PW and EW are less likely. We infer that the system is viewed at an intermediate viewing angle $i\sim\ang{35}$, and we estimate the most likely \ion{C}{iv} BLR half-light radius $r_\mathrm{1/2}=51\pm\SI{23}{}$ light days. Our results are in good agreement with previous findings in the literature and extend the validity of the index-based approach to a temporal domain. 

}


\maketitle

\section{Introduction}
Gravitational lenses are important phenomena used to studying cosmology and galaxy formation and evolution. The magnification of distant faint objects allows us to constrain the mass of the foreground lenses based on the fluxes and the position of the images. Owing to their intrinsic variability \citep{1997ARA&A..35..445U,10261_216313}, lensed quasars have been turned into powerful cosmographic probes by means of the measurement of time delays between the different images \citep{1964MNRAS.128..295R,2022arXiv221010833B}.


Extragalactic gravitational microlensing is a natural phenomenon that occurs when the light emitted by a quasar is bent and focused by the gravity of a single star in the foreground galaxy, causing a temporary increase in the quasar's brightness \citep{1979Natur.282..561C,2023arXiv231200931V}. When detected, microlensing can be exploited to constrain the parameters of the quasar emitting regions \citep{1988ApJ...335..593N,1992ApJ...396L..65J,2000MNRAS.315...62W,2002ApJ...576..640A,2008ApJ...673...34P,2010ApJ...712..668P,2011ApJ...729...34B,2012A&A...544A..62S,2016ApJ...830..149F}.

Type 1 quasars are characterized by prominent broad emission lines (BELs) in their optical spectra \citep{1993ARA&A..31..473A,2012A&AT...27..557A,2017A&ARv..25....2P}. The BELs are emitted from the   broad-line region (BLR), which  consists of ionized gas situated in the vicinity of the supermassive black hole \citep[SMBH,][]{2013peag.book.....N}. In a lensed quasar, the BLR projected size is typically a few times larger than the microlensing Einstein radius of the deflecting stars. For that reason, subregions of the BLR may be differently affected by microlensing such that deformation of the emission line profiles are   observed \citep{2003A&A...398..975P,2005MNRAS.357..135P,2004MNRAS.348...24L,2004ApJ...610..679R,2008MNRAS.386..397J,2007A&A...468..885S,2011A&A...528A.100S,2012A&A...544A..62S,2013ApJ...764..160G,2014A&A...565L..11B,2016A&A...592A..23B,2017ApJ...835..132M,2020A&A...634A..27P,2018ApJ...859...50F,2021A&A...653A.109F,2023A&A...677A..94F,2023arXiv231011212F,2023A&A...678A.108F,2019A&A...629A..43H,2023A&A...672A..45H,2021A&A...654A.155H}. 

The majority of quasars remain spatially unresolved. Although interferometry is able to probe the subparsec-scale regions in the most luminous and closest objects \citep{2018Natur.563..657G}, such observations will remain impossible for the vast majority of quasars that are observed at high redshifts. The alternative method commonly used to probe the BLR structure, velocity resolved reverberation mapping (RM) \citep{1972ApJ...171..467B,1982ApJ...255..419B,1993PASP..105..247P,2000ApJ...533..631K,2021ApJ...915..129K}, becomes considerably more telescope time intensive for active galactic nuclei (AGNs) at redshift $z>1$. Microlensing nicely complements other methods  as it is independent of the intrinsic variability of the source.

One of the key difficulties that arise when microlensing is applied in the BLR studies is that the microlensing events are rare and unpredictable, and typically last for several years \citep{2011ApJ...738...96M}. As a result,  it is challenging to gather multi-epoch spectroscopic data covering a complete event. Single-band photometric data, obtained for time delay cosmography, have now been obtained for several tens of systems \citep[e.g.,~COSMOGRAIL,\footnote{\url{https://www.epfl.ch/labs/lastro/scientific-activities/cosmograil/}}][]{2020A&A...640A.105M}, but multi-epoch spectrophotometry is rarer.

Recently, \citet{2017A&A...607A..32B} have proposed a method to constrain the BLR structure based on the study of microlensing-induced line deformations. For that purpose, they   characterize the effect of microlensing through four measurable quantities: $\mu^{cont}$, the magnification of the continuum underlining the emission line; $\mu^{BLR}$, the total magnification of the broad emission line; and  $WCI$ and $RBI$, the indices sensitive to wing-core and red-blue line profile distortions. \citet{2019A&A...629A..43H}  developed a probabilistic Bayesian framework through which we are  able to constrain the geometry, inclination, and effective size of the BLR by comparing those quantities to similar ones derived from simulated microlensed line profiles.


The methodology introduced by \citet{2017A&A...607A..32B} used indices as measured at a single epoch, however, it is yet unclear if this strategy is free of biases when studying a single object. Potential biases on microlensing size inference from the single epoch analyses have been suggested for accretion disk temperature profile studies \citep{2018MNRAS.479.4796B}. While these results may not be directly applicable to BLR analyses that simultaneously constrain the accretion disk and the BLR size, it is desirable to investigate whether the modeling of the BLR with multiple epochs measurements agrees with single-epoch modeling. In addition, the time variation of microlensing is equivalent to a scan of the BLR, potentially providing fine-grained constraints on the BLR structure enabling us to break degeneracies between BLR models. For those reasons, we expanded the previous modeling scheme to the analysis of multiple epochs of the same system.

As a first test case of this new framework, we selected a well-studied quadruply lensed system Q2237$+$0305. The past observations of images A-D    revealed that during the monitoring period, image D remained basically unaffected by microlensing while image A was subject to high magnification \citep{2008A&A...480..647E}. In particular, we used archival data to study the geometry and size of the \ion{C}{iv} emitting region. 

In section \ref{sec:obs} we describe the observations and explain how the indices used for the microlensing analysis are derived. Section \ref{sec:mlm} describes our microlensing models, and section \ref{sec:res} our main results. Finally, section \ref{sec:conc} summarizes our key findings and lists our conclusions.

\section{Observational data}
\label{sec:obs}
The gravitational lens system Q2237$+$0305 (also known as Einstein cross or Huchra's lens) consists of four quasar images of similar brightness that are separated by $\ang{;;1.6}$. The quasar is at redshift $z_{\rm s}=1.695$ and the lensing galaxy is at $z_{\rm l} = 0.0394$ \citep{1985AJ.....90..691H}. The A/D macro-magnification ratio is $M=\SI{1.0(1)}{}$ \citep{2000ApJ...545..657A}. The time delay between the lensed images is negligible ($<1$ day), such that any difference between pairs of images can be attributed to microlensing.

The system was spectrophotometrically monitored with the FORS1\footnote{\url{https://www.eso.org/sci/facilities/paranal/instruments/fors.html}} instrument
(ESO Very Large Telescope) in the multi-object spectroscopy (MOS) observing mode from October 2004 to December 2007. Data reduction and calibration were reported by \citet{2008A&A...480..647E} and \citet{2011A&A...528A.100S}, and will not be repeated here.

The analysis of the spectrophotometric data of the four lensed images of Q2237$+$0305 presented in \citet{2008A&A...480..647E} supports the absence of microlensing in image D during the monitoring. While image B was also   found to be minimally affected by microlensing over that period, its use as a reference for the analysis was considered   sub-optimal. The spectra of the lensed images were observed in pairs, with A-D and B-C obtained in consecutive observations. While this observational strategy is optimal for slit spectroscopy, it resulted in a different number of usable spectra for images A and B,   sometimes with substantial differences in data quality for spectra obtained at the same epoch.

\begin{figure}[ht!]
    \centering
    \includegraphics[width=0.99\hsize]{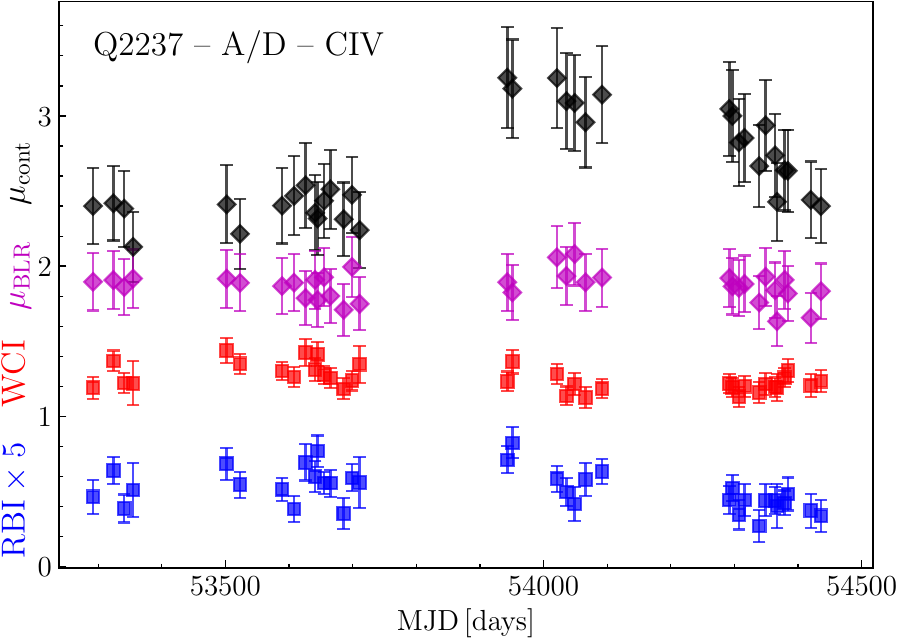}
    \caption{Time series of the four indices computed for \ion{C}{iv} line from the spectra of images A and D obtained at the same epoch. The indices $\mu^{cont}$, $\mu^{BLR}$, $WCI$, and $RBI$ are denoted in black, magenta, red, and blue, respectively. The values for $RBI$ are multiplied by 5 for clarity.}
    \label{fig:Q2237CIV.pdf}
\end{figure}

Following \citet{2017A&A...607A..32B}, we reduce the microlensing signal to the measurement of four quantities: $\mu^{cont}$, $\mu^{BLR}$, $WCI$, and $RBI$. The microlensing-induced magnification factors of the continuum $\mu^{cont}$ were estimated at the wavelength of the \ion{C}{iv} line from the adjacent A/D continuum ratios, corrected for the differential extinction and macro-magnification. We briefly recall the definition of the  $\mu^{BLR}$, WCI, and RBI indices  \citep{2019A&A...629A..43H}: 
\begin{equation}
    \mu^{BLR} = \frac{1}{M}\dfrac{\int_{v_-}^{v_+}F^l_A(v)dv}{\int_{v_-}^{v_+}F^l_D(v)dv}, \label{eq:mublr}
\end{equation}
\begin{equation}
    WCI = \dfrac{\int_{v_-}^{v_+}\mu(v)/\mu(v=0)dv}{\int_{v_-}^{v_+}dv}, \label{eq:muwci}
\end{equation}
\begin{equation}
    RBI = \dfrac{\int_0^{v_+}\log \mu(v)dv}{\int_0^{v_+}dv} - 
    \dfrac{\int_{v_-}^0\log \mu(v)dv}{\int_{v_-}^0dv}, \label{eq:murbi}
\end{equation}
\begin{equation}
    \mu(v)=\dfrac{F^l_A(v)}{M\times F^l_D(v)}.
\end{equation}
\noindent Here $F^l_A$ and $F^l_D$ are continuum-subtracted flux densities in the emission lines, corrected for the differential extinction and the A/D macro-magnification ratio $M$. The limits $v_-,\,v_+$ are integration boundaries over a restricted velocity range. The $RBI$ index is sensitive to the asymmetry of the line deformations. It takes non-null values when the amplitude of microlensing is different in the blue and red parts of the line. The $WCI$ index indicates whether the wings of the emission line are affected by microlensing with respect to the line core. Both RBI and WCI are independent of $M$. As a flux ratio, $\mu(v)$ can be extremely noisy in the wings of the emission lines where the flux density reaches zero, so it is necessary to cut the faintest parts of the line wings.  We thus only considered the parts of the line profiles whose flux density is above $l_{\rm cut} \times F_{\rm peak}$, where $F_{\rm peak}$ is the maximum flux in the line profile and $l_{\rm cut}$ is fixed to 0.1 at all epochs. This value is a compromise between a good signal-to-noise ratio in the line wings and the preservation of a large part of the line profile. The values of the observed indices for the various epochs are shown in Fig.\ \ref{fig:Q2237CIV.pdf}. The continuum magnification strongly increased between MJD~53711 and MJD~53943 suggesting a caustic crossing event. On the other hand, the three indices $\mu^{BLR}$, $WCI$, and $RBI$ do not show significant variations, suggesting that the line profile deformations, while strong \citep[][see also Appendix~\ref{sec:appendix}]{2021A&A...654A.155H}, remain essentially constant during the monitoring period. The indices measurements are given in Table \ref{tab:measurements}.

\begin{table}[t]
    \centering
    \caption{Dates of observations (MJD, first column) and measured indices $\mu^{cont}$, $\mu^{BLR}$, $WCI$, $RBI$ with errors respectively.}
    \label{tab:measurements}
    \resizebox{\columnwidth}{!}{%
    \begin{tabular}{ccccccccc}
    \hline
    \hline
    MJD & $\mu^{cont}$ & $\Delta\mu^{cont}$ & $\mu^{BLR}$ & $\Delta\mu^{BLR}$ & $WCI$ & $\Delta WCI$ &
    $RBI$ & $\Delta RBI$ \\
    \hline
    53292 & 2.40 & 0.25 & 1.90 & 0.19 & 1.19 & 0.07 & 0.09 & 0.02 \\ 
    53324 & 2.42 & 0.25 & 1.91 & 0.19 & 1.37 & 0.07 & 0.13 & 0.02 \\ 
    53341 & 2.38 & 0.25 & 1.86 & 0.19 & 1.22 & 0.07 & 0.08 & 0.02 \\ 
    53355 & 2.13 & 0.23 & 1.92 & 0.19 & 1.22 & 0.15 & 0.10 & 0.04 \\ 
    53502 & 2.41 & 0.26 & 1.92 & 0.19 & 1.44 & 0.08 & 0.14 & 0.02 \\ 
    53523 & 2.22 & 0.23 & 1.89 & 0.19 & 1.35 & 0.07 & 0.11 & 0.02 \\ 
    53589 & 2.40 & 0.25 & 1.87 & 0.19 & 1.30 & 0.06 & 0.10 & 0.02 \\ 
    53608 & 2.47 & 0.26 & 1.89 & 0.19 & 1.26 & 0.07 & 0.08 & 0.02 \\ 
    53626 & 2.54 & 0.28 & 1.79 & 0.18 & 1.42 & 0.09 & 0.14 & 0.02 \\ 
    53641 & 2.36 & 0.25 & 1.91 & 0.19 & 1.31 & 0.07 & 0.12 & 0.02 \\ 
    53645 & 2.32 & 0.24 & 1.77 & 0.18 & 1.41 & 0.08 & 0.15 & 0.02 \\ 
    53655 & 2.44 & 0.25 & 1.93 & 0.19 & 1.28 & 0.06 & 0.11 & 0.01 \\ 
    53665 & 2.51 & 0.26 & 1.80 & 0.18 & 1.26 & 0.07 & 0.11 & 0.02 \\ 
    53686 & 2.31 & 0.24 & 1.71 & 0.17 & 1.18 & 0.07 & 0.07 & 0.02 \\ 
    53699 & 2.47 & 0.25 & 2.00 & 0.20 & 1.24 & 0.07 & 0.12 & 0.02 \\ 
    53711 & 2.24 & 0.25 & 1.75 & 0.18 & 1.35 & 0.12 & 0.11 & 0.03 \\ 
    53943 & 3.25 & 0.34 & 1.89 & 0.19 & 1.23 & 0.07 & 0.14 & 0.02 \\ 
    53951 & 3.18 & 0.33 & 1.83 & 0.18 & 1.37 & 0.08 & 0.17 & 0.02 \\ 
    54021 & 3.25 & 0.33 & 2.06 & 0.21 & 1.28 & 0.07 & 0.12 & 0.02 \\ 
    54036 & 3.10 & 0.32 & 1.94 & 0.19 & 1.14 & 0.06 & 0.10 & 0.02 \\ 
    54049 & 3.09 & 0.32 & 2.08 & 0.21 & 1.22 & 0.07 & 0.08 & 0.02 \\ 
    54066 & 2.96 & 0.30 & 1.89 & 0.19 & 1.13 & 0.07 & 0.12 & 0.02 \\ 
    54092 & 3.14 & 0.32 & 1.92 & 0.19 & 1.18 & 0.06 & 0.13 & 0.02 \\ 
    54292 & 3.05 & 0.31 & 1.92 & 0.19 & 1.22 & 0.07 & 0.09 & 0.02 \\ 
    54297 & 3.00 & 0.31 & 1.87 & 0.19 & 1.19 & 0.06 & 0.11 & 0.02 \\ 
    54307 & 2.82 & 0.29 & 1.86 & 0.19 & 1.13 & 0.07 & 0.07 & 0.02 \\ 
    54316 & 2.85 & 0.29 & 1.88 & 0.19 & 1.20 & 0.07 & 0.09 & 0.02 \\ 
    54339 & 2.67 & 0.28 & 1.76 & 0.18 & 1.16 & 0.07 & 0.05 & 0.02 \\ 
    54349 & 2.94 & 0.30 & 1.93 & 0.19 & 1.22 & 0.07 & 0.09 & 0.02 \\ 
    54364 & 2.74 & 0.28 & 1.84 & 0.18 & 1.20 & 0.06 & 0.09 & 0.02 \\ 
    54367 & 2.43 & 0.26 & 1.63 & 0.17 & 1.19 & 0.09 & 0.08 & 0.03 \\ 
    54379 & 2.64 & 0.27 & 1.91 & 0.19 & 1.26 & 0.06 & 0.09 & 0.02 \\ 
    54384 & 2.63 & 0.27 & 1.82 & 0.18 & 1.31 & 0.08 & 0.10 & 0.02 \\ 
    54420 & 2.44 & 0.26 & 1.66 & 0.17 & 1.20 & 0.08 & 0.07 & 0.02 \\ 
    54436 & 2.40 & 0.25 & 1.83 & 0.18 & 1.23 & 0.07 & 0.07 & 0.02 \\ 
    \hline
    \end{tabular}}
\end{table}

The uncertainties on the indices are obtained by propagating the uncertainty of the line flux densities. This uncertainty is computed as the quadratic sum of the uncertainty on the total (line + continuum) flux density and the uncertainty on the continuum estimate, the latter being taken as the standard deviation of the continuum flux on each side of the emission lines. The exact value of the indices depends on several parameters, such as the adopted continuum windows, the velocity range fixed by $l_{\rm cut}$, and the systemic redshift that defines the line center. Considering different, reasonable, values of these parameters, we estimated the additional uncertainty around 0.05 for $WCI$  and $0.01$  for $RBI$. The uncertainties are added quadratically. The uncertainty of  $\mu^{cont}$ and $\mu^{BLR}$, on the other hand, are dominated by the uncertainty that affects the macro-magnification factor $M$.

\section{Microlensing model}
\label{sec:mlm}
In order to infer the properties of the source, a forward-modeling method is generally followed that simulates microlensing data and compares them to the observations \citep{2023arXiv231200931V}. Microlensing simulations include two parts: 1) the convolution of the source projected image, accretion disk, or BLR models, with the magnification map that represents the caustic network, and 2) the linear motion of the source over the convolved magnification map for a given transversal velocity in an arbitrary direction. I think you mean this: In this case, a magnification event is considered successful if the values along an arbitrary linear path of a fixed length over each of the convolved images reproduce the observed indices time series within the error bars. Our approach is similar to the approximate Bayesian computation \citep[ABC,][]{2013PLSCB...9E2803S}: the four indices are our summary statistics and are compared to similar summary statistics from the data to evaluate the model's likelihood. In the following subsections, we describe in detail the modeling steps.


\subsection{Cosmological parameters and microlensing map}
We assumed a flat universe with cosmological parameters: $H_0 = \SI{68}{km\,s^{-1}\,Mpc^{-1}}$, $\Omega_m = 0.31$, $\Omega_\Lambda = 0.69$ \citep{2021A&A...654A.155H}. The associated angular diameter distances for the system are $D_{\mathrm{ol}} = \SI{166}{Mpc}$ (observer-lens), $D_{\mathrm{os}} = \SI{1793}{Mpc}$ (observer-source), and $D_{\mathrm{ls}} = \SI{1729}{Mpc}$ (lens-source). Due to the absence of microlensing in image D at the time of the observations, we  assumed that the whole microlensing signal was associated with image A. 

The microlensing map was computed using the code \textsc{microlens}\footnote{\url{https://github.com/psaha/microlens}} \citep{1999JCoAM.109..353W}. We considered a convergence $\kappa_s = 0.394$ for matter in compact objects; $\kappa_c = 0$ for continuously distributed matter due to the fact that the lensed images lie behind the bulge of the lens such that the dark matter fraction towards the lensed images is zero; and shear $\gamma=0.395$ \citep{2004ApJ...605...58K}. We set the mean microlens mass to $\left \langle m \right \rangle = \SI{0.3}{\ensuremath{\mathcal{M}_\odot}}$. The microlensing map is normalized such that the mean magnification is 1. The microlensing Einstein radius projected in the source plane is 
\begin{equation}
    r_\mathrm{E} = D_{\mathrm{os}} \sqrt{\dfrac{4G\langle m \rangle}{c^2}\dfrac{D_{\mathrm{ls}}}{D_{\mathrm{ol}}D_{\mathrm{os}}}} = \SI{39}{\ensuremath{ld}},
\end{equation}
where $G$ is gravitational constant and $c$ is the speed of light. The total size of the magnification map, when projected to the source plane, is $\SI{200}{\ensuremath{r_\mathrm{E}}}\times\SI{200}{\ensuremath{r_\mathrm{E}}}$ sampled with a resolution of $\SI{20000}{}\times \SI{20000}{\ensuremath{pixel}}$. To reduce the impact of the preferred direction, as well as the edge effects, the caustic map was rotated by $\theta = [0, 30, 45, 60, 90^\circ]$ and only the central part of $\SI{10000}{}\times \SI{10000}{\ensuremath{pixel}}$ was used for analysis.

\subsection{Continuum and broad-line region parameters}
We followed the same model setup as used in \citet{2017A&A...607A..32B}. A continuum-emitting uniform disk is situated in the center surrounded by the BLR. A total of nine different outer radii of the continuum source are used: $r_\mathrm{s} = [0.1, 0.15, 0.2, 0.25, 0.3, 0.4, 0.5, 0.6, 0.7]\SI{}{\ensuremath{r_\mathrm{E}}}$. We used three different geometries of the BLR: Keplerian disk (KD), polar wind (PW), and equatorial wind (EW). Both PW and EW are radially accelerated. The BLR emissivity depends on the radius in the form of a power law with index $q=[1.5, 3.0]$. The BLR inner radius is in the range $r_\mathrm{in} = [0.075, 0.1, 0.125, 0.15, 0.175, 0.2, 0.25, 0.35, 0.5, 0.75]\SI{}{\ensuremath{r_\mathrm{E}}}$. The outer radius of the BLR source is ten times larger than the inner size. The whole system is viewed at four different inclinations $i = [22, 34, 44, 62]^\circ$ with respect to the polar axis. A system viewed at $i = \ang{0}$ corresponds to a face-on view. For a full description of the model setup, we refer to \citet{2017A&A...607A..32B} and \citet{2019A&A...629A..43H}. The emitted continuum and BLR images are computed using the radiative code \textsc{stokes}\footnote{\url{http://www.stokes-program.info/}} \citep{2007A&A...465..129G,2012A&A...548A.121M,2015A&A...577A..66M,2018A&A...615A.171M,2018A&A...611A..39R}, which  is publicly available. All the models have the same number of degrees of freedom.

\subsection{Extracting index time series}
In order to simulate caustic crossing events that reproduce the observed indices, we convolved in the source plane the magnification map with each monochromatic image of the BLR obtained by slicing the projected BLR image in 20 velocity bins, thus obtaining a data cube of 20 convolved maps. From this cube, we computed the indices $\mu^{BLR}$, $WCI$, and $RBI$ (Eqs. \ref{eq:mublr}, \ref{eq:muwci}, and \ref{eq:murbi}), which left us with three maps. For the continuum emission, we only had one monochromatic image. Convolved with the magnification map, it constitutes the fourth map. To emulate the  time series to be compared to the data, it was necessary to extract tracks from the four maps \citep{2004ApJ...605...58K,2011A&A...528A.100S}. The conversion of the track length into a time length was done through the net transverse velocity of the system. In the simplest case, the projected transverse velocity in the source plane $v_\perp(\mathrm{source\,plane})$ for a randomly oriented track, reduces to \citep{1986A&A...166...36K,2004A&A...416...19T,2012MNRAS.425.1576S} 
\begin{equation}
  v_\perp(\mathrm{source\,plane}) = \frac{D_{\mathrm{os}}}{D_{\mathrm{ol}}}v_\perp(\mathrm{lens\,plane}).
\end{equation}
For this work, we considered four different values of the transverse velocity $v_\perp=[300, 400, 500, 600]\SI{}{\kilo\meter\per\second}$ in the lens plane, as found and used by several studies of Q2237+0305 \citep{1990ApJ...352..407W,1990ApJ...358L..33W,1999MNRAS.309..261W,2004ApJ...605...58K,2005A&A...432...83G,2010ApJ...712..658P,2011ApJ...738...96M,2020MNRAS.495..544N}. The velocities projected in the source plane were about ten times larger than in the lens plane. 

\begin{figure*}[h]
    \centering
    \includegraphics[width=0.99\hsize]{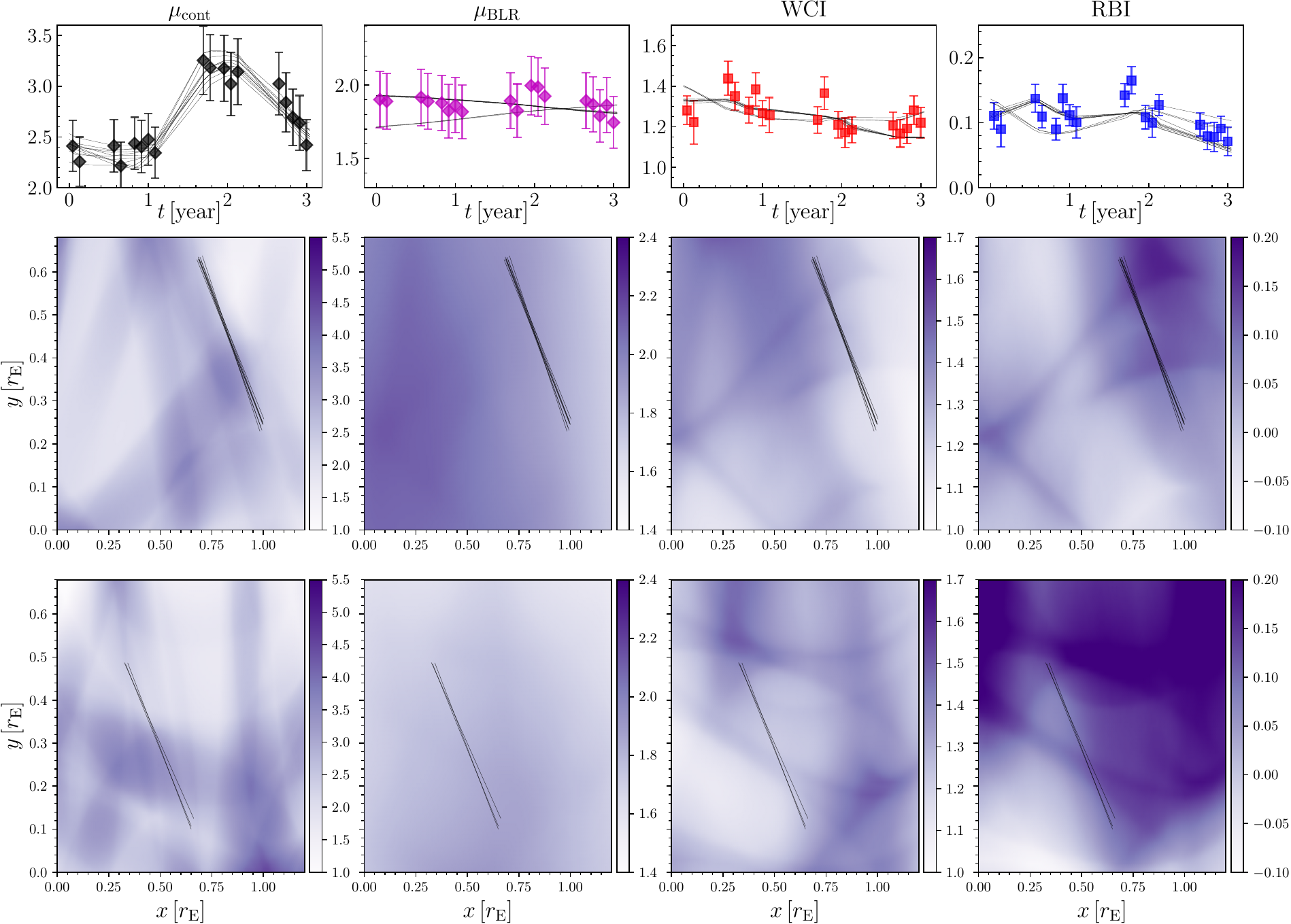}
    \caption{Microlensing model used to reproduce the observed index time series. From left to right: $\mu^{cont}$, $\mu^{BLR}$, $WCI$, and $RBI$. Top panels: Rebinned time series of the four observed indices computed from the spectra of images A and D obtained at the same epoch. The solid lines denote the reconstructed time series. Bottom panels: Tracks (solid lines) sampled from two different regions that reproduce the indices in the top panels. In that example, the BLR model is characterized as KD, $i=\ang{22}$, $q=1.5$, $r_\mathrm{in}=\SI{0.5}{\ensuremath{r_\mathrm{E}}}$, $r_\mathrm{s}=\SI{0.1}{\ensuremath{r_\mathrm{E}}}$. The bottom panels present only two out of many regions on a much wider map. The units of the tracks are converted to $r_\mathrm{E}$ for $v_\perp = \SI{500}{\kilo\meter\per\second}$.}
    \label{fig:indiceslines.pdf}
\end{figure*}

Track extraction was performed using a Monte Carlo approach based on uniform sampling. First, we smoothed the observed index time series by rebinning the signal into sparser time intervals in order to mitigate the variability on the smallest timescales that cannot be spatially resolved on the magnification map (Fig.\,\ref{fig:indiceslines.pdf}, top panels). For a given track on the convolved map, the index time series were obtained by interpolating the values along the track (Fig.\,\ref{fig:indiceslines.pdf}, bottom and middle panels). The formalism introduced by \citet{2019A&A...629A..43H} for computing the likelihood of the observed indices for each set of parameters characterizing the simulations was simply extended to time domain (i.e.,\,the definition of a microlensing event was extended from a point to a track). We performed the extracting procedure for all convolved maps and the whole parameter space of the models and we computed the likelihoods, which were used to reweight the samples (importance sampling). The total number of sampled tracks is on the order of $\sim\SI{5d8}{}$ per map.

\section{Results}
\label{sec:res}
We obtained the relative probabilities of the BLR models for each value of the inclination by marginalizing the likelihood over other parameters (Table \ref{tab:tsresults}). The EW models were almost totally rejected. The KD models are the most likely, while PW models are less probable overall, confirming the results of \citet{2021A&A...654A.155H} based on single-epoch data. We estimate the most likely viewing inclination of the system $i=\SI{35}{}\pm\ang{12}$, in agreement with the independent analysis by \citep{2010ApJ...712..668P}.
\begin{figure}[ht]
    \centering
    \includegraphics[width=0.99\hsize]{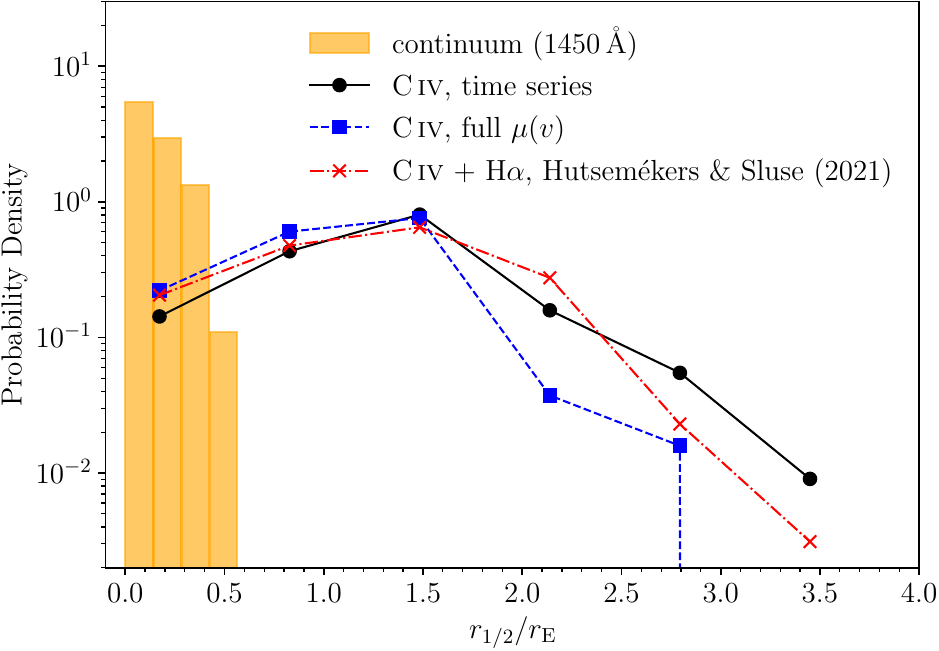}
    \caption{Posterior probability distributions of the continuum and the BLR half radius. The solid black line denotes measurements using index time series for \ion{C}{iv} line. The dashed blue line denotes the measurements using the full $\mu(v)$ profile at a single epoch. Red dot-dashed line denotes a joint probability reported by \citet{2021A&A...654A.155H} using the \ion{C}{iv} and H$\alpha$ lines.}
    \label{fig:Rhalf.pdf}
\end{figure}

As stated by \citet{2021A&A...654A.155H}, $r_\mathrm{in}$ does not properly represent the BLR size due to strong dependence on inclination and preferred geometry. We followed the same prescription for computing the values of the BLR half radii $r_\mathrm{1/2}$ \citep{2021A&A...654A.155H} and relative probabilities of $r_\mathrm{1/2}$ (Fig.\,\ref{fig:Rhalf.pdf}). From the probability distribution, we obtain $r_\mathrm{1/2}=1.31\pm0.60\,r_\mathrm{E}=51\pm\SI{23}{\ensuremath{ld}}$, in agreement with values reported by \citet{2021A&A...654A.155H}, or from the same time series but modeling the total microlensing amplitude of \ion{C}{iv} reported by \citet{2011A&A...528A.100S}.

We estimate the effective continuum half size $r_\mathrm{s} = \SI{0.16\pm0.11}{\ensuremath{r_\mathrm{E}}} = \SI{6.0\pm4.1}{\ensuremath{ld}}$, in accordance with earlier estimates \citep{2000MNRAS.315...62W,2004ApJ...605...58K,2008A&A...480..327A,2008A&A...480..647E,2010ApJ...712..668P,2011A&A...528A.100S,2016ApJ...817..155M,2016ApJ...831...43V}.

\begin{table}[t]
    \centering
    \caption{Relative probabilities (in \%) for BLR models marginalized over emissivity power law indices $q$ and BLR sizes using \ion{C}{iv} indices time series.}
    \label{tab:tsresults}
    \begin{tabular}{ccccc}
    \hline
    \hline
    \multicolumn{5}{c}{\ion{C}{iv} time series} \\
    \hline
             & KD & PW & EW & ALL\\
    \hline
    22\degr  & 30 & 1 & 1 & 32\\
    34\degr  & 20 & 9 & 1 & 30\\ 
    44\degr  & 24 & 4 & 2 & 30\\
    62\degr  & 7  & 1 & 0 & 8\\
    All $i$  & 81 & 15 & 4 \\
    \hline
    \end{tabular}
\end{table}

Recently, \citet{2023A&A...672A..45H} used the full $\mu(v)$ magnification profile instead of the indices (Sect. \ref{sec:obs}) in   comparison to simulations, to constrain the size, geometry, and kinematics of the BLR in the lensed quasar J1004+4112, based on single-epoch data. They found that using either the indices or the full $\mu(v)$ profile gave similar results. In Appendix~\ref{sec:appendix} we report similar results for Q2237+0305, using the full $\mu(v)$ profile at a single epoch. This is also illustrated in Fig.\,\ref{fig:Rhalf.pdf}, thus validating the use of integrated indices, which are more convenient to handle when analyzing time series. When compared to single-epoch estimates based on isolated points on the maps that reproduce observations, sampling time series acts as a filter to spatially uncorrelated signals along an arbitrary path.



Recently, \citet{2023A&A...677A..94F,2023arXiv231011212F} investigated a sample of 13 quadruply lensed quasars in order to study the influence of diffuse BLR emission \citep{2001ApJ...553..695K,2019MNRAS.489.5284K,2022MNRAS.509.2637N} on the accretion disk size inferences using microlensing. They showed that the mere contribution of the BLR to the continuum signal is able to largely account for the implied accretion disk overestimation, and that microlensing may provide useful constraints on disk physics in sources whose diffuse BLR emission is weak and extends much farther than the typical Einstein radius, such as in the most luminous sources. Although Q2237+0305 is not included in their sample, our simulations imply that the BLR effective size for this object is much larger than the continuum region that radiates at \SI{1450}{\ensuremath{\mathrm{\mbox{\AA}}}}; however, a certain scenario ($\sim \SI{15}{\percent}$ of total cases) where the outer part of the disk overlaps with the inner funnel of the BLR is possible, but less likely, and could indicate a weak diffuse BLR emission. The high values observed for $\mu^{cont}$ are statistically favored by compact continuum sources in our model (i.e.,\,the continuum source and the BLR are geometrically distinct and rarely overlap). 

\section{Conclusions}
\label{sec:conc}
Current and future generations of wide-field surveys such as GAIA \citep{2016A&A...595A...1G}, LSST \citep[Vera C.~Rubin Observatory Legacy Survey of Space and Time,][]{2019ApJ...873..111I}, and  EUCLID \citep{2022A&A...662A.112E} are expected to identify thousands of bright lensed quasars that are able to be robustly modeled \citep{2010MNRAS.405.2579O,2023MNRAS.tmp.2164T}. For each successful detection of a caustic crossing event, an immediate ground-based spectroscopy follow-up may play a key role in understanding the structure and evolution of quasars at high cosmological redshifts.

In order to develop a new framework for the upcoming stream of high-cadence time series data, we studied the microlensing effect of the \ion{C}{iv} emission line observed for the quadruply lensed system Q2237+0305 during a monitoring performed between October 2004 and December 2007. We simulated realistic caustic crossing events as a linear motion of a point source over convolved source images in order to reproduce the time series of the four representative indices $\mu^{cont}$, $\mu^{BLR}$, $WCI$ and $RBI$ that are capable of characterizing a single microlensing effect. We   explored a wide range of quasar internal parameters in order to determine the most likely geometry, size, and orientation of the continuum and line emitting source. Based on our simulations, we conclude the following:
\begin{itemize}
    \item We confirm that the index analysis developed by \citet{2017A&A...607A..32B} and \citet{2019A&A...629A..43H} is valid when applied to time domain and allows for maximum usage of spectroscopic monitoring data for probing the BLR structure. 
    \item The most likely geometry for Q2237+0305 is KD, while PW and EW are less likely. The effective \ion{C}{iv} emitting BLR size we inferred is in good agreement with the previous measurements \citep{2021A&A...654A.155H}. 
\end{itemize}

Several improvements of the presented method will be included in a follow-up research. Microlensing simulations reproducing the observed signal in more than two lensed images would allow us to narrow the probability distribution on the continuum source and the BLR size with the goal of using the time series of the full $\mu(v)$ differential magnification profile that is the most sensitive to geometry. It will be addressed in the future work. We note, however, that using full $\mu(v)$ time series requires considerably higher computational resources, due to the dimensionality increase. Future prospects will also include the application of the same method to other broad emission lines observed for quadruply lensed systems, for example  SDSS J1004+4112 and  RX J1131-1231.

\begin{acknowledgements}
We thank the anonymous referee for valuable comments that improved the quality of the manuscript. This work was supported by the F.R.S.\,FNRS under the research grants IISN 4.4503.19 and PDR T.0116.21.
\end{acknowledgements}

\bibliography{bibliography}
\bibliographystyle{apj}

\begin{appendix}

\section{Single-epoch results using either the indices or the full $\mu(v)$ profile}
\label{sec:appendix}

\begin{figure}[!b]
\centering
\resizebox{0.95\hsize}{!}{\includegraphics*{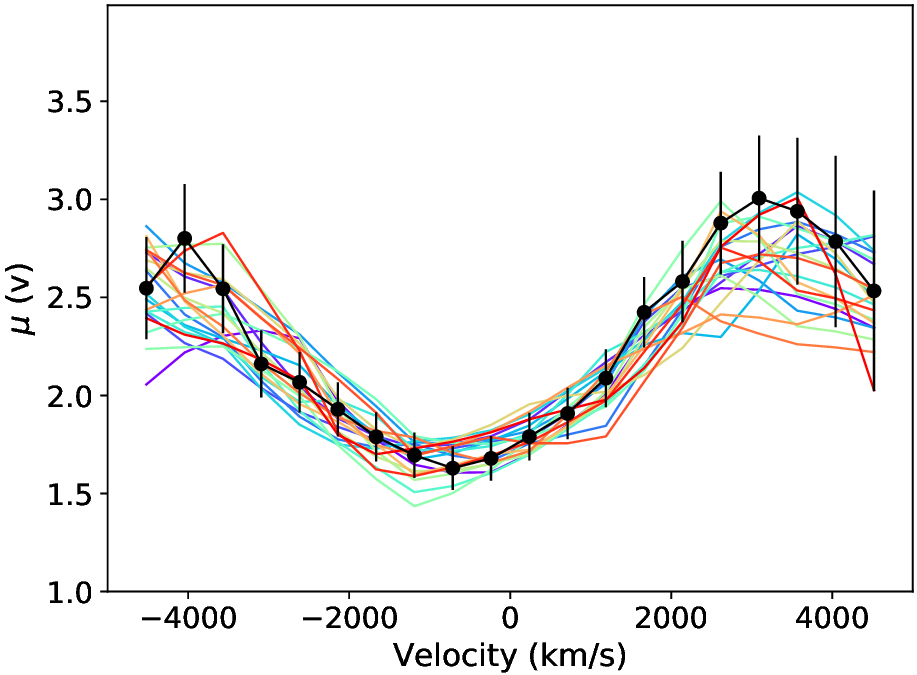}}
\caption{Example of 20 simulated $\mu(v)$ profiles (in color) that fit the $\mu(v)$ profiles measured for the \ion{C}{iv} emission line (in black with error bars). The simulated profiles were computed for the KD model with $i = 34\degr$, $q = 3$, $r_{\rm in} = 0.1 \, r_{\rm E}$, and selected with $\chi^{2} / n_{\rm d.o.f.} \leq 1.2$.}
\label{fig:fitmuvQ2237civ}
\end{figure}

\begin{figure}[!b]
\centering
\resizebox{0.95\hsize}{!}{\includegraphics*{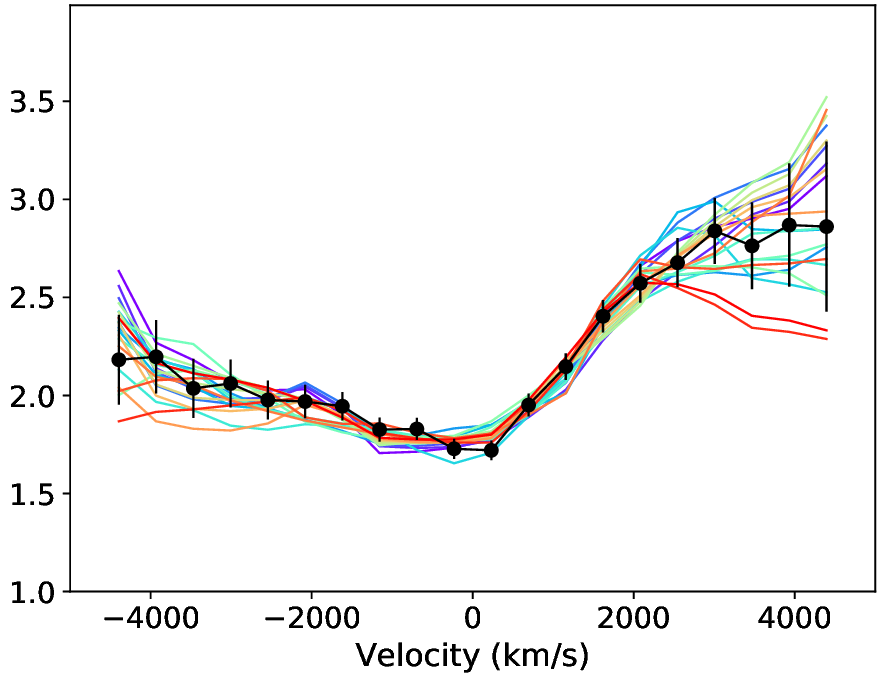}}
\caption{Same as Fig. \ref{fig:fitmuvQ2237civ}, but  for the H$\alpha$ emission line.}
\label{fig:fitmuvQ2237ha}
\end{figure}

\citet{2021A&A...654A.155H}  analyzed the microlensing-induced line profile deformations observed quasi-simultaneously in the \ion{C}{iv} and H$\alpha$ line profiles, at a single epoch in October 2005, to constrain the size, geometry, and kinematics of the BLR in Q2237+0305. This study was done using the four indices  $\mu^{cont}$, $\mu^{BLR}$, $WCI$, and $RBI$ to characterize the microlensing effect. Hereafter, we update this study considering the $\mu(v)$ magnification profile in the comparison to simulations, as recently done for the lensed quasar J1004+4112 \citep{2023A&A...672A..45H}. For consistency, the three indices characterizing the emission line microlensing are recomputed from line profiles slightly truncated to discard the noisiest parts of the  $\mu(v)$ profile (in October 2005; the data were good enough to use $l_{\rm cut}$ = 0.06). These indices are given in Table~\ref{tab:indices}. For \ion{C}{iv}, the values of  $\mu(v)$, $\mu^{cont}$, and $\mu^{BLR}$ are 25\% smaller due to the differential extinction in image D which was not taken into account in \citet{2017A&A...607A..32B} and \citet{2021A&A...654A.155H} (for H$\alpha$, the values of  $\mu^{cont}$ and $\mu^{BLR}$ are only 3\% smaller). Details of the method can be found in \citet{2021A&A...654A.155H}.

As shown in Figs.~\ref{fig:fitmuvQ2237civ} and ~\ref{fig:fitmuvQ2237ha}, the observed $\mu(v)$ profiles can be reproduced by many simulated profiles. The probabilities of the different BLR models are given in Table~\ref{tab:probaQ2237}, for \ion{C}{iv} and H$\alpha$. We find the following. First, the probabilities derived from the four updated indices are in good agreement with those reported in \citet{2021A&A...654A.155H}, for both \ion{C}{iv} and H$\alpha$. Second, the probabilities derived using the 20 spectral elements of the $\mu(v)$ profile are in good agreement with those obtained from the indices, for both \ion{C}{iv} and H$\alpha$. Third, compared to previous results, the probability of the EW model slightly increases, but the KD model remains the dominant one. Finally, the measurement of the half-light radius of the BLR, based on the probabilities computed using the full $\mu(v)$ profiles of \ion{C}{iv} and H$\alpha$ separately, gives $r_{1/2} = 39^{+17}_{-25}$ light-days for \ion{C}{iv} and $r_{1/2} = 37^{+12}_{-23}$  light-days for H$\alpha$, in excellent agreement with the values reported in \citet{2021A&A...654A.155H}.

\begin{table}
\caption{New magnification and distortion indices}
\label{tab:indices}
\centering
\begin{tabular}{p{0.03\textwidth}>{\centering}p{0.085\textwidth}>{\centering}p{0.085\textwidth}>{\centering}p{0.085\textwidth}>{\centering\arraybackslash}p{0.085\textwidth}}
\hline\hline
  Line  & $\mu^{cont}$ & $\mu^{BLR}$ & $WCI$ & $RBI$ \\
\hline
\ion{C}{iv}         & 2.45$\pm$0.25 & 1.95$\pm$0.19 & 1.31$\pm$0.04 & 0.089$\pm$0.014 \\
H$\alpha$           & 2.38$\pm$0.24 & 2.03$\pm$0.20 & 1.28$\pm$0.05 & 0.093$\pm$0.011 \\
\hline
\end{tabular}
\phantom{X} \\
\phantom{X} 
\caption{Probability (in \%) of BLR models for \ion{C}{iv} and H$\alpha$}
\label{tab:probaQ2237}
\centering
\begin{tabular}{p{0.06\textwidth}>{\centering}p{0.024\textwidth}>{\centering}p{0.024\textwidth}>{\centering}p{0.024\textwidth}>{\centering}p{0.024\textwidth}p{0.01\textwidth}>{\centering}p{0.024\textwidth}>{\centering}p{0.024\textwidth}>{\centering}p{0.024\textwidth}>{\centering\arraybackslash}p{0.024\textwidth}}
\hline\hline
     & \multicolumn{4}{c}{\ion{C}{iv}} &  & \multicolumn{4}{c}{H$\alpha$} \\
\hline
     \multicolumn{10}{c}{Using the 4 indices} \\
\hline
          & KD & PW &EW & ALL & & KD & PW & EW & ALL \\
\hline
          22\degr         & 22 &  1 &  0 & 24   & & 22 &  2 &  1 & 24 \\
          34\degr         & 23 &  5 &  2 & 30   & & 22 &  4 &  3 & 28 \\
          44\degr         & 22 &  3 &  3 & 28   & & 22 &  2 &  3 & 27 \\
          62\degr         & 15 &  2 &  3 & 19   & & 16 &  2 &  3 & 21 \\
          All $i$         & 82 & 11 &  7 &      & & 82 &  9 &  9 &    \\
\hline
     \multicolumn{10}{c}{Using $\mu^{cont}$ and the $\mu(v)$ profile} \\
\hline
          & KD & PW & EW & ALL & & KD & PW & EW & ALL \\ 
\hline
          22\degr         & 31 &  0 &  0 & 31   & & 34 &  0 &  0 & 34 \\
          34\degr         & 20 &  1 &  1 & 22   & & 19 &  0 & 13 & 32 \\
          44\degr         & 17 & 15 &  2 & 34   & & 14 &  2 &  7 & 22 \\
          62\degr         & 11 &  1 &  1 & 13   & & 10 &  0 &  2 & 12 \\
          All $i$         & 79 & 17 &  5 &      & & 76 &  2 & 22 &    \\
\hline
\end{tabular}
\end{table}

The results are thus robust with respect to small changes in the indices. Moreover, using the full $\mu(v)$ profile does not modify the results, including the determination of the BLR size. This validates the use of integrated indices in simulations of microlensing-induced line profile deformations, at least for Q2237+0305 and J1004+4112.

\end{appendix}

\end{document}